\documentclass[letterpaper,twocolumn,10pt]{article}
\usepackage{usenix2019_v3}

\usepackage{tikz}
\usepackage{amsmath}
\usepackage{filecontents}
\usepackage{color}
\usepackage{listings}
\begin{document}

\title{\Large \bf S3Library: Automatically Eliminating C/C++ Buffer Overflow 
	\\using Compatible Safer Libraries}

\author{
	{\rm Kang Sun, Daliang Xu, Dongwei Chen, Xu Cheng, Dong Tong}\\
	Department of Computer Science and Technology, Peking University\\
	\{ajksunkang, xudaliang, chendongwei, tongdong\}@pku.edu.cn, chengxu@mprc.pku.edu.cn
} 
	
\maketitle
	
\begin{abstract}
	
Annex K of C11, bounds-checking interfaces, recently introduced a set of alternative functions to mitigate buffer overflows, primarily those caused by string/memory functions. However, poor compatibility limits their adoption. Failure oblivious computing can eliminate the possibility that an attacker can exploit memory errors to corrupt the address space and significantly increase the availability of systems.
		
In this paper, we present S3Library (Saturation-Memory-Access Safer String Library), which is compatible with the standard C library in terms of function signature. Our technique automatically replaces unsafe deprecated memory/string functions with safer versions that perform bounds checking and eliminate buffer overflows via boundless memory. S3Library employs MinFat, a very compact pointer representation following the $Less$ $is$ $More$ principle, to encode metadata into unused upper bits within pointers. In addition, S3Library utilizes Saturation Memory Access to eliminate illegal memory accesses into boundless padding area. Even if an out-of-bounds access is made, the faulty program will not be interrupted. We implement our scheme within the LLVM framework on X86-64 and evaluate our approach on correctness, security, runtime performance and availability.  

\end{abstract}

\section{Introduction}
	
Buffer overflows remain a major threat to the security
of dependable computer systems today \cite{szekeres2013sok}. Applications written in low-level languages like assembly or C/C++ are prone to buffer overflow bugs. From 2008 to 2014, nearly 23 percent of all severe software vulnerabilities were buffer errors, and 72 percent of buffer errors were serious \cite{homaei2017seven}. In 2018, among the 16,556 security vulnerabilities recorded by the NIST National Vulnerability Database \cite{NIST}, 2,368 (14.3\%) were overflow vulnerabilities. Meanwhile, buffer overflow is listed as the first position in Weaknesses CWE Top 25(2019) \cite{cwetop25}.
	
The danger inherent in the use of unsafe standard C library calls, especially the string/memory functions, presents classic buffer overflows. These contiguous overflows still cover over 40 percent in real-world exploitation \cite{twitter}. Early designers of standard C (specifically ANSI) over-trusted the programmers. A host of vulnerable functions calls they provide, such as $gets$, $strcpy$ and $memcpy$, neither make bounds checking to determine whether the destination buffer is big enough nor have the size information needed to perform such checks.
	
Annex K of C11, Bounds-checking interfaces \cite{jtc2011sc22, N997, TR24731}, recently proposed a set of new, optional alternative library functions that promote safer, more secure programming. The apparent difference (we will explain in later section) is that the APIs have $\_s$ suffix and take an additional $size$ argument explicitly passed by programmers. Intuitively, adopting these APIs in an existing code requires non-trivial modifications leading to poor compatibility and guidance. This is the main reason why the new APIs continue to be controversial \cite{N1106,seacordbounds, N1967}, despite almost a decade since its introduction. Furthermore, there is almost no viable conforming implementation \cite{laverdiere2009implementation} applying the bounds-checking interfaces without considerable origin code changes.
	
Various modern techniques have been proposed to enforce memory safety, both statically and dynamically. Static analysis~\cite{monperrus2018automatic, shahriar2013buffer, smirnov2007automatic, shaw2014automatically} that automatically transforming C programs at source code level is hard to obtain a complete coverage because a certain type of size information is only available at runtime. 
Dynamic defense mechanisms \cite{akritidis2009baggy,  nagarakatte2009softbound, serebryany2012addresssanitizer, duck2017stack, kuvaiskii2017sgxbounds, oleksenko2017intel, kroes2018delta} augment the original unmodified program with metadata (bounds of live objects or allowed memory regions) and insert bounds checking against this metadata before every memory access for runtime detection. They all leave libraries uninstrumented and introduce manually written wrapper to maintain the compatibility, performing simple bounds checking before calling a real legacy function. However, all existing software-based bounds-checking solutions exhibit high performance overhead (50-150\%), preventing them from wide adoption in production runs. Address Sanitizer \cite{serebryany2012addresssanitizer} is currently better in terms of usability, but it is built for debugging purposes and suffers from detecting non-contiguous out-of-bounds violations. Intel MPX \cite{oleksenko2017intel} provides a promising hardware-assisted full-stack technique, but its implementation is proved not as good enough as expected. Most of these approaches provide complete protection for buffer overflow violations, detecting both contiguous and non-contiguous overflows, but have relatively high runtime overhead. 
	
Performing bounds checking is costly due to large amounts of metadata management: the metadata describing the object bounds must be recorded, propagated and retrieved to check numerous times. Among these processes, the checking is the bottleneck. For each pointer dereference, metadata must be loaded from memory/in-pointer to verify the validation of the pointer. Once the pointer is out of bounds, it also gives much pressure on the branch predictor and pipeline to handle exceptions. As a result, the checking process accounts for the vast majority of execution time.
	
In this paper, we propose an interesting idea in the exploration space to concentrate only on buffer overflows caused by highly-critical memory/string functions, rather than bounds checking on each memory reference. With this ``partial'' memory safety, we wonder what trade-off we can achieve between security and overhead. Meanwhile, we propose a feasible implementation of Safe C Library without any modification to existing C programs.
	
We present MinFat and S3Library, an interesting approach to automatically replace unsafe deprecated functions like $strcpy$ with safer versions that perform bounds checking and eliminate buffer overflows via boundless memory. MinFat is based on the tagged pointer \cite{kroes2017fast, serebryany2018memory}
scheme that transparently encodes bounds meta information of
buffer (stack, heap and global variables) within the pointer itself. We follow the principle of $Less\ is\ More$ and adopts a very compact encoding scheme inspired by BaggyBounds~\cite{akritidis2009baggy} with the minimum bit-width that allows an effective way to retrieve object bounds. 
S3Library retains the APIs compatibility with legacy functions, and performs the same bounds checking as Safe C Library. The property of MinFat trades memory for performance and adds boundless padding \cite{rinard2004dynamic,rinard2004enhancing,rinard2003acceptability,brunink2011boundless} to every object. These boundless memory blocks support Saturation Memory Access (SMA) to isolate the memory errors within S3Library in case the runtime-constraint violation occurs.


Overall, this paper makes the following contributions:
\begin{itemize}
	\item To the best of our knowledge, S3Library is the first runtime solution that applies implementations of Safe C Library without any modification of origin codes.
	\item A thorough analysis and evaluation of the overhead using MinFat Pointer on selective functions.
	\item We present a buffer overflow elimination mechanism within Safe C Library and evaluate its performance in Section~\ref{runtime overhead}. 
	\item An LLVM-based prototype of our design implemented in X86-64 architecture environment, evaluated with respect to security, availability and runtime performance.
\end{itemize}

	

\section{Background}
	
\begin{figure}[t]
	\setlength{\abovecaptionskip}{0pt}
	\begin{center}
		\includegraphics[width=0.5\textwidth]{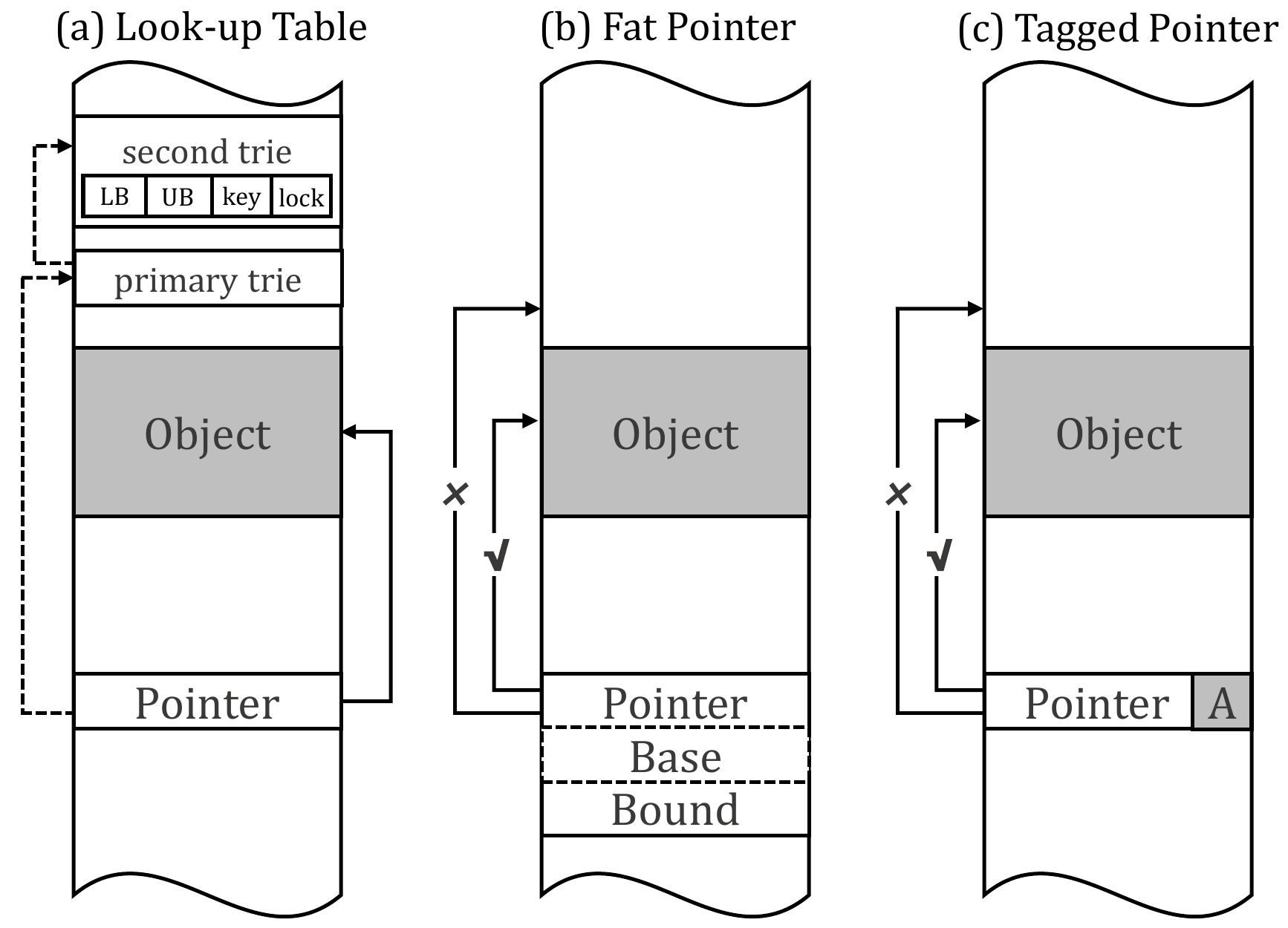}	
	\end{center}
	\caption{\label{background-crop} Pointer-based approach.}
\end{figure}
	
\subsection{Memory Safety}
	
Enforcing $Memory$ $Safety$ stops all memory corruption exploits. Existing runtime techniques that guarantee spatial and temporal memory safety can be broadly categorized into two classes: object-based approach and pointer-based approach. Object-based approaches~\cite{serebryany2012addresssanitizer, ruwase2004practical,jones1997backwards} associate metadata with the location of the object in memory, not with each pointer. The significant drawback is that its implementations are generally incomplete because they are unable to provide an accurate bound information for each object. However, an accurate size information is necessary for the implementation of Safe C Library to perform bounds checks. Besides, object-based approaches are not suitable for non-contiguous buffer overflow detection also due to the lack of accurate bounds information.

	
In this section, we focus on eliminating spatial memory errors (i.e., buffer overflows) using the pointer-based approach, which is considered as the only way to support comprehensive memory safety~\cite{nagarakatte2015everything}.
Pointer-based approaches ~\cite{akritidis2009baggy,  nagarakatte2009softbound,  duck2017stack, kuvaiskii2017sgxbounds, oleksenko2017intel, kroes2018delta, woodruff2014cheri} attach metadata with every pointer, bounding the region of memory that pointer can legitimately dereference. As presented in Figure~\ref{background-crop}, pointer-based approaches can be categorized into the following three classes according to how current designs store and use metadata.
	
\noindent\textbf{Look-up Table scheme: SoftBound}~\cite{nagarakatte2009softbound,nagarakatte2010cets}. This class, such as SoftBound, records base and bound metadata in a disjoint metadata facility that is accessed by explicit table look-ups. Figure~\ref{background-crop}(a) shows the way how SoftBound+CETS organizes pointer metadata in a two-level trie for quick searching purpose. With table look-ups, the accurate bound information can be obtained in time whenever the pointer needs. Unfortunately, this look-up table scheme cannot be considered safe in multithread environments. Ideally, the load/store of the pointer and its bounds must be performed atomically. However, neither the current hardware implementations (e.g., Intel MPX\cite{oleksenko2017intel}) nor GCC/LLVM compilers enforce this atomicity. Furthermore, this look-up table scheme imposes restrictions on allowed memory layout, and some programs require substantial code changes to run correctly.

\noindent\textbf{Fat Pointer scheme: CHERI}~\cite{woodruff2014cheri}.
One implementation of pointer-based metadata is to expand all pointer into multi-word $fat$  $pointers$~\cite{xu2004efficient}. For example, CHERI uses 128-bit fat pointers/capabilities to limit the range of memory that each pointer is allowed to access, as presented in Figure~\ref{background-crop}(b). The base and bound metadata fields always follow behind the pointer and are maintained with a pointer structure. The fat pointers give the system extra accuracy information needed to call Safe C Library to avoid buffer overflow, but also change the space requirement for a pointer. Consequently,  compatibility with the precompiled library or kernel is lost. The modified pointer representation makes it challenging to interface with external libraries due to its ABI breaking nature.

\noindent\textbf{Tagged Pointer scheme: Low-Fat Pointer}~\cite{duck2017stack,kwon2013low}. Tagged Pointer scheme is a new method for tracking bounds information that takes advantage of 64-bit systems with sufficient pointer bit-width. The basic idea is to store the boundary meta information within the representation of the machine pointer itself, as shown in Figure~\ref{background-crop}(c). Pointer tagging avoids memory layout changes that the look-up table scheme results in. Another benefit is that it can load the pointer and meta information in only one memory operation, bringing no pressure on the cache while retrieving the metadata. Besides, tagged pointer has the same pointer representation as standard C in contrast with fat pointer scheme, resulting in a better ABI compatibility.


\subsection{Safe C Library}
	
Annex K of C11, bounds-checking interfaces, was intended to introduce a set of new, optional alternative functions into the standard C library to mitigate the security implication of a subset of buffer overflows in existing code. Historically, the design of Annex K dates back to an ISO/IEC technical report in 2007 before being incorporated in C11 as normative Annex K~\cite{jtc2011sc22, N997, TR24731}.

However, the software incompatibility~\cite{szekeres2013sok} makes these interfaces hard to adopt, despite over a decade since the introduction of them~\cite{N1106, N1967}. Specifically, these bounds-checking interfaces have apparent differences in the API. These APIs have $\_s$ suffix, commonly take an additional argument and return a value of type $errno\_t$ rather than a pointer to the destination buffer. For example, legacy $strcat()$ is declared as follows:
	

\begin{lstlisting}
char* strcat(char* dest, const char* src);
\end{lstlisting}
	
\noindent It corresponds to the Annex K $strcat\_s$ (the syntactic difference is in bold font) whose declaration is the following:
	
\begin{lstlisting}
`\textbf{errno\_t}` strcat`\textbf{\_s}`(
  	char* dest,
  	`\textbf{rsize\_t dmax,}`
  	const char* src);
\end{lstlisting}
	
Usability is critically essential in the design of a secure interface. Unfortunately, adopting the APIs in millions of lines of existing C/C++ legacy code requires non-trivial modification mainly due to the following reasons\cite{seacordbounds, N1967}:

\noindent\textbf{Extra size argument.} The additional argument $dmax$ (the size of destination buffer) is not always readily available at the site of the replacement. While the length of the pointer to an array can be determined using static analysis, a certain type of storage information dynamically allocated is only available at runtime.
If we only pass the parameter using the $sizeof$ operator, the overhead is apparent when the same destination buffer is used as an argument to multiple API calls in the same function. Besides, it can easily result in some programming errors when the programmer incorrectly specifies the size of source buffer ($sizeof$ src) instead of the destination array.
	
\noindent\textbf{Distinct return value.} The Annex K supports additional error handling features. It deploys the uniform return value of $errno\_t$ to indicate the status of the returning functions, i.e., NULL pointer and zero length. While $errno\_t$ provides a general exception handling mechanism, however, it will lead to mistakes for the case when multiple API calls are in a single expression. For example:

\begin{lstlisting}
  strcat ( strcpy(d, a), b);
\end{lstlisting}	

\noindent the return value of strcpy records the result and the strcat uses the return value of strcpy as the first parameter. The single expression must be rewritten as two separated statements:
	
\begin{lstlisting}
  strcpy_s ( d, sizeof d, a );
  strcat_s ( d, sizeof d, b );
\end{lstlisting}
	
\noindent\textbf{Error handling policy.} Annex K, C11 introduces a new term into the C
standard, namely runtime-constraint violations. When a function detects an error (such as invalid parameter or reference outside the bounds), a specific function (i.e., runtime-constraint handler) is called that either abort the program or just issue an error message. This is in sharp contrast to the runtime error handling in the standard C library,
where the behavior under such errors is mostly undefined (anything may happen then, like buffer overflow). For security-critical systems where forcing the system to terminate may indicate the goal of an attack, continued execution\cite{rinard2004dynamic,rinard2004enhancing,rinard2003acceptability} results in security compromise. 
For example, embedded devices must provide robust and continuous service in the presence of unexpected inputs. However, the current term runtime-constraint handler lacks an elimination (or tolerant) way to make computing invulnerable to known security attacks without interrupting the normal execution path.

For the sake of convenience, we refer to all implementation of bounds-checking interfaces as $Safe$ $C$ $Library$. As a consequence of API incompatibility, experience with these functions has not been supported well by the mainstream compiler~\cite{openwatcom, SLIBC, SafeCLibrary, IntelSafeString}. The largest body of experience in implementation comes from Microsoft. Unfortunately, Microsoft implementation conforms neither to C11 nor to the original TR 24731-1~\cite{seacordbounds}.
In short, there is almost no viable conforming implementation applying Safe C Library without considerable source code changes. This motivates our case for an implementation of Safe C Library with the same API compatibility as legacy functions.

\section{Design}
\subsection{Threat Model}

We assume that an attacker can exploit the victim program to gain arbitrary read and write capabilities in the memory. Temporal memory errors, e.g., use-after-free vulnerability, are not considered in this paper. Furthermore, we assume that all processor hardware can be trusted-it does not contain vulnerabilities arising from exploits such as physical or glitching attacks. Our goal is to prevent spatial memory violations caused by highly-critical unsafe legacy functions. Once an out-of-bound memory access occurs, the faulty program will not be interrupted. The un-contiguous buffer overflow that happens in non-library codes is not under our protection.

\subsection{MinFat Pointer}
	
Our MinFat is a tagged pointer scheme that takes advantage of 64-bit modern systems with sufficient pointer bit-width. The basic idea is to encode, propagate and retrieve bounds metadata information within tag bits of the pointer while object allocation, pointer arithmetic and bounds check. 

\begin{figure}[ht]
	\setlength{\abovecaptionskip}{0pt}
	\begin{center}
		\includegraphics[width=0.49\textwidth]{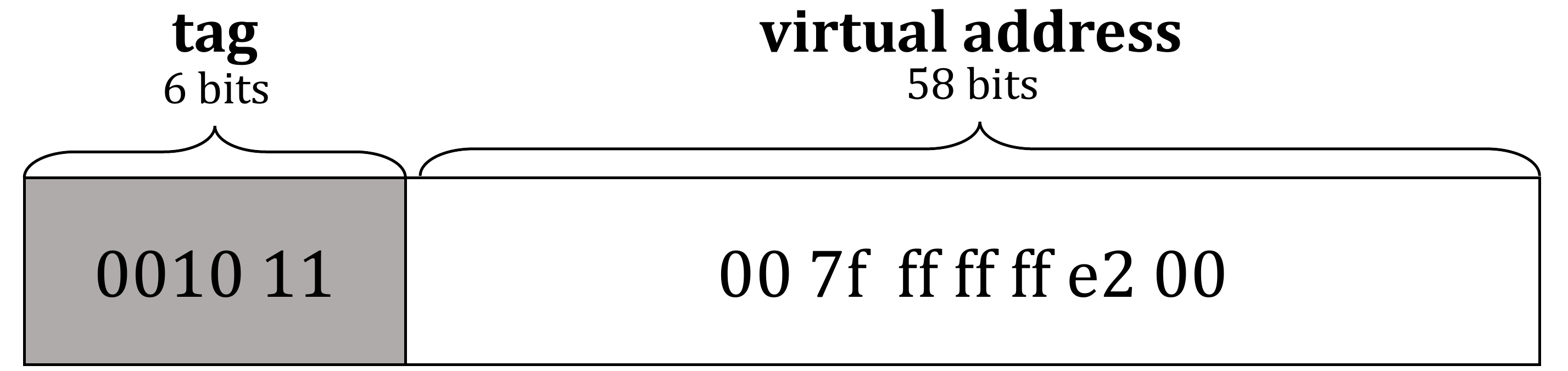}	
	\end{center}
	\caption{\label{tag-pointer} The encoding scheme of MinFat Pointer. The upper 6 bits represent the allocated size of the memory object. The remaining 58 bits are used for virtual address.}
\end{figure}

\noindent\textbf{Pointer Encoding.} We extend the pointer tagging scheme with a compact encoding format, perhaps the $minimum$ one along with the ability to cover almost all objects in memory. This is the reason why we name it MinFat Pointer. Specifically, MinFat Pointer uses the upper six bits to encode the object size, allowing for a maximum allocation size of $2^{64}$ bytes. The remaining 58 bits are used to address in 64-bit virtual memory space.

Each of the MinFat Pointer is associated with a base address ($base$ for short) and specific allocation size ($allocSize$ for short). It is noted that the allocation size of an object is always powers-of-two. The size configuration:
	
$$allocSize =< 2, 4, 8, 16, 32, ...,2^{64} > Bytes$$
	
\noindent specifies that the MinFat allocator supports allocation sizes of 2 Bytes, 4 Bytes, 8 Bytes, 16 Bytes, etc. The maximum allocation size is $2^{64}$ bytes, which is large enough to hold any memory object in a 64-bit modern system. 
The MinFat allocator ensures that every buffer or object should be aligned to $allocSize$. Thus we can reconstruct the $base$ using the following operations:
$$base(p) = (p\ /\ allocSize(p))\ *\ allocSize(p)$$
Correspondingly, we can also reconstruct the $allocSize$ and the TAG value recorded in upper bits as follows:
$$allocSize(p)\ =\ (p\ \&\ MINFAT\_MATCH)\ >>\ 58$$
$$TAG = \log_2(allocSize(p))$$
	
Here (\&), ($*$) and ($/$) are 64-bit integer bitwise AND, division, and multiplication. $MINFAT\_MATCH$ is the matching value 0x03ffffffffffffff. Figure~\ref{tag-pointer} illustrates our encoding scheme: the upper 6 bits are MinFat TAG, followed by 58 bits virtual address. We record value B in the tag bits. The allocation size of an object will be $2^{B}$ bytes. Thus, in Figure~\ref{tag-pointer}, the tag $0010 11$ (whose decimal value is 11) represents the $2^{11}=2048$ bytes allocation size for this object.
	
Our MinFat encoding scheme comes from the “buddy system” method~\cite{akritidis2009baggy} of dynamic memory allocation algorithm, which expands the object size of an allocation to its nearest power-of-2. In contrast to prior schemes, the key insight exploited by our encoding scheme is that MinFat occupies the minimum bit-width within the pointer by default (but configurable) and also can effectively represent large enough object size in various real-world scenarios.
	
\begin{figure}[t]
	\setlength{\abovecaptionskip}{0pt}
	\begin{center}
		\includegraphics[width=0.49\textwidth]{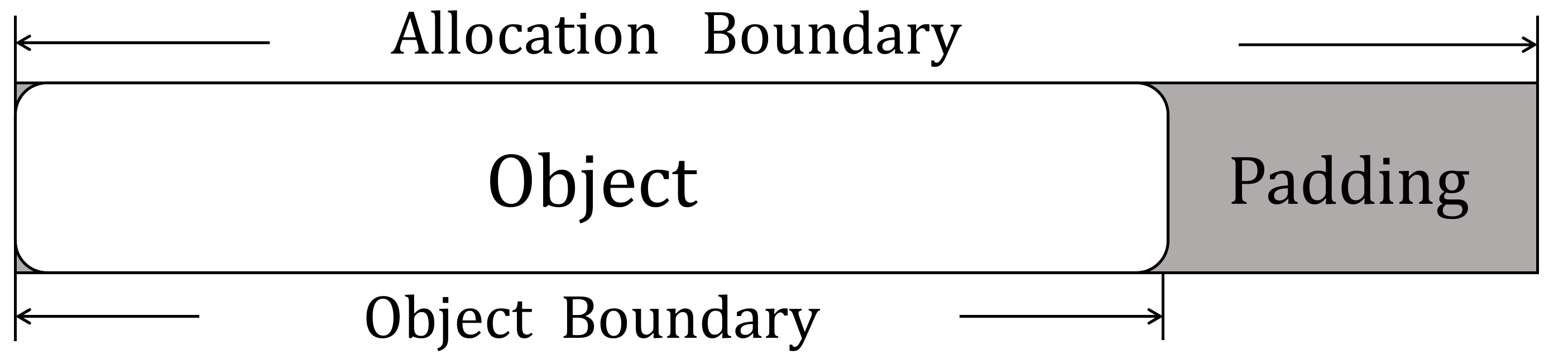}	
	\end{center}
	\caption{\label{allocator} Aligned and Padding in MinFat. The allocation memory is always padded to a pow-of-2 alignment boundary.}
\end{figure}
	
\noindent\textbf{Padding and Aligning.} The memory object in C/C++ comes from three areas:
stack, heap and global variables. During the allocation of these
objects, the requested object size (i.e., the size parameter to
malloc) is aligned (over-approximated) to the nearest allocation
size $allocSize$ that fits. Figure~\ref{allocator} presents the padding and aligning rationale in MinFat Pointer. For example, consider an object $O$ of type $char[20]$ with an object size of
$sizeof(O) = 20$. Assuming the above size configuration, the size will be rounded up to an allocation size of $allocSize = 32$ by adding 12 bytes of padding.

To provide enough space for the defense mechanism against out-of-bounds memory access, at least one byte of padding is
always added (so $allocSize$ > object size) to ensure our object will not be violated. These padding bytes are usually zero values, allowing for benign access by out-of-bounds memory operation and we will discuss in Section~\ref{sec:S3Library}. The boundary checking is not to check the requested size, but the allocation size recorded in the tag bits. Although MinFat allocation algorithm results in some memory fragments, its effective encoding scheme makes it easy to extract the boundary of a memory object as quickly as possible.

\noindent\textbf{Pointer Arithmetic.} In our threat model, we ignore the case when pointers reference outside the bounds of an array or buffer in application code.
In this case, when an expression contains pointer arithmetic(e.g., $ptr$+$index$) or pointer assignment(e.g., $newptr$ =$ptr$), the $baseptr$ and $allocSize$ is not going to be affected. Thus, the result pointer inherits the same TAG of the original pointer because they belong to the same object and has the same $allocSize$. As for pointer comparison, we remove their tags for a fair comparison.
	
\noindent\textbf{Pointer Dereference.} The X86\_64 standard require that the upper bits of pointer value are sign-extended. Therefore, any MinFat pointer dereference in application code has to trigger the masking process to create the regular untagged pointer the CPU expects. Since our scheme is not devoted to detect buffer overflows in non-library code, we do not perform boundary checking before each MinFat pointer dereference.

\noindent\textbf{Function Calls.} For selected string and memory functions, MinFat pointers are passed directly as parameters to library calls where they are masked at the entry of each function. After the dereferencing and bounds-checking operations, the tags are restored back at the end of the function call (if returns a pointer), which offers protection for the next use. For other external functions, we introduce manually written wrapper and make masking operating in the wrapper functions before calling a real libc function.
	
\subsection{S3Library}
\label{sec:S3Library}

\begin{figure}[t]
	\setlength{\abovecaptionskip}{0pt}
	\begin{center}
		\includegraphics[width=0.49\textwidth]{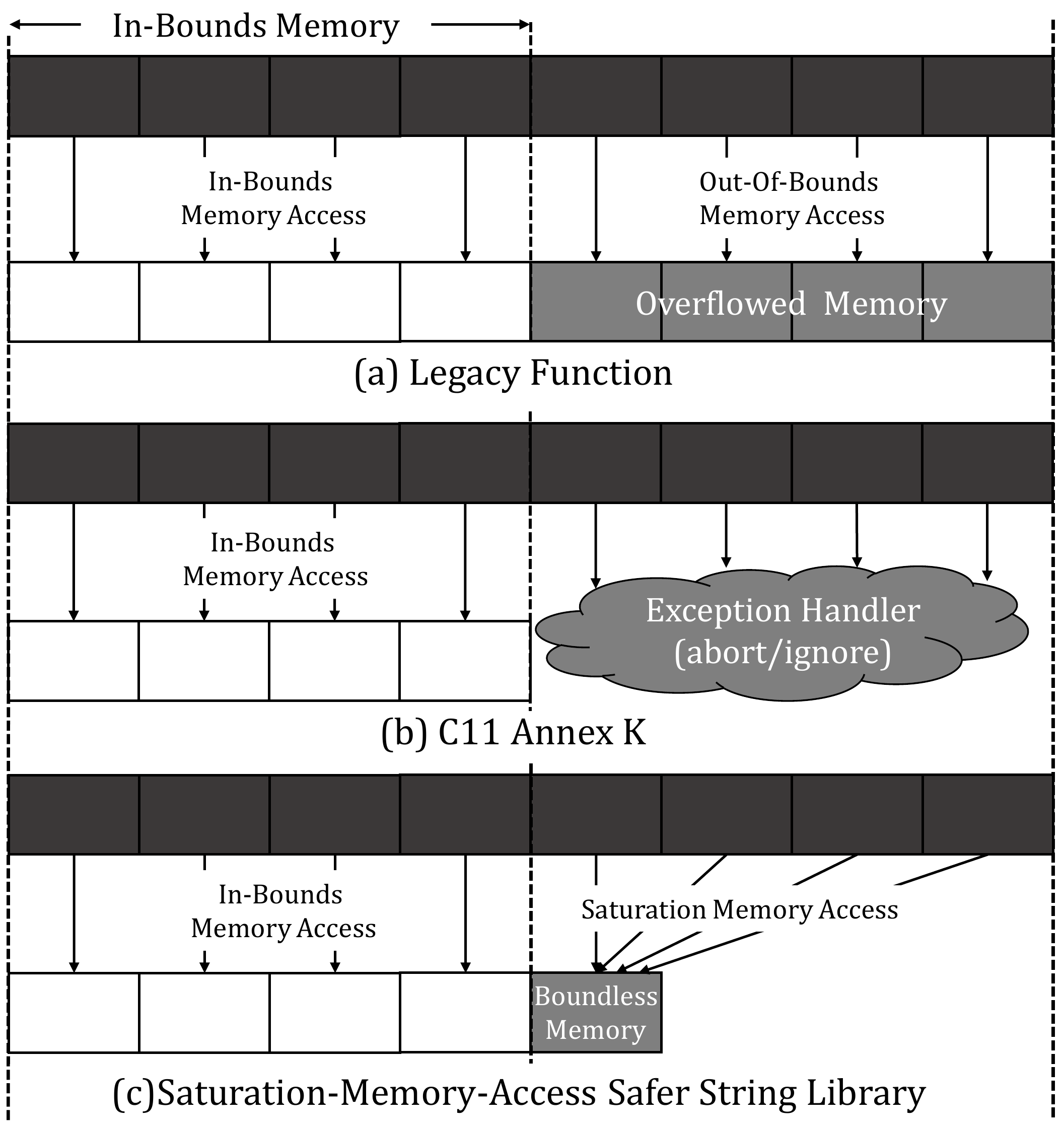}	
	\end{center}
	\caption{\label{SMA} Error handling policy of legacy function (a), C11 Annex K (b) and S3Library (c).}
\end{figure}

\noindent\textbf{Elimination and Detection.}
Existing defenses mechanisms handle buffer overflow using a detection-only, fail-stop approach, e.g.,~\cite{akritidis2009baggy,  nagarakatte2009softbound}. Annex K of C11, bounds-checking interfaces also call the $abort\_handler\_s$ for terminating the function call in case a runtime-constraint violation occurs. However, terminating
a program (or function) allows potential Denial-of-Service (DoS) attacks~\cite{rinard2004enhancing}, and decreases the availability of the system. In security-critical systems (such as an airplane that must provide robust and continuous opetations in case of various unexpected inputs) forcing the system to terminate may indicate the goal of an attack, and continued execution\cite{rinard2004dynamic,rinard2004enhancing,rinard2003acceptability} results in security compromise. 
For example, servers and embedded devices all need continued running to service legitimate requests. 

Instead of terminating, systems can choose an elimination way that manufactures a boundless~\cite{brunink2011boundless} value within the allocation boundary as the result of the illegal memory access and keeps the program running with that value. The elimination approach can achieve the same goal of isolating memory errors without violating the data integrity property. Besides, the error handling policy of Safe C Library also lacks an elimination way to make the library's implementation invulnerable to known security attacks without terminating the programs.

\noindent\textbf{Saturation Memory Access.}
In this section, we propose $Saturation$ $Memory$ $Access$ (SMA) to provide an elimination way for the implementation of Safe C Library. MinFat provides $boundless$ $memory$ $blocks$ to out of bounds accesses. Instead of allowing out of bounds accesses to corrupt the address space or aborting the program by an exception, SMA utilizes these boundless memory blocks and takes actions that allow the program continued execution without interruption.

As presented in Figure~\ref{allocator}, MinFat adds additional padding bytes to every object so that its size will be power of two and base address will be calculated by aligning the allocation size. The property not only allows an effective way to lookup object bounds but also automatically adds the tolerant feature to C programs.

We derived our elimination (or tolerant) way from failure oblivious computing~\cite{rinard2004enhancing}. 
The implementation of failure oblivious computing simply discards the illegal memory writes, and out-of-bounds reads values are forged based on a heuristic. In SMA, instead of discarding and forging, out-of-bounds memory accesses are redirected into the last padding bytes. These padding bytes allows benign access by out-of-bounds memory operations and referencing them does not violate the data integrity property. SMA manufactures the boundless memory value within the allocation boundary as the result of illegal memory access to keep the programs running with that value.

\noindent\textbf{S3Library.} There are two major barriers to the adoption of Safe C Library. One barrier is software incompatibility. Adopting the new APIs in existing codes requires non-trivial modification. In general, a safe C library function needs programmers to provide an extra size parameter, either explicitly or implicitly, to perform checks within the function.
The problem of explicit passing is that original APIs must be rewritten and source code compatibility will be lost. In our design, MinFat allows implicitly pass way of the size parameter into a safe C library function without API modifications.

The other barrier is its incomplete error handling policy. The current term runtime-constraint handler lacks an elimination way to automatically isolate memory errors without interruption. In our design, SMA enables the computing to continue its normal execution path through memory errors.

In order to break the two barriers, we propose Saturation-memory-access Safer String Library ($S3Library$) to combine the feature of MinFat and Saturation Memory Access to reinforce the implementation of Safe C Library. Specifically, the encoded metadata within MinFat Pointer is implicitly passed into the library function to perform bounds checking, allowing S3Library retains the same API as legacy functions. Meanwhile, S3Library performs SMA to eliminate the runtime-constraint violation by correcting the illegal address into a redundant memory block and the faulty program will not be terminated. Continued execution can significantly increase the availability of S3Library.

Figure~\ref{SMA} presents the difference in principle between S3Library and Annex K bounds-checking interfaces over buffer overflow. Take a string-copy function as an example. In our case, the behavior of legacy functions is mostly undefined and dangerous memory operations continue until the program crashes, as shown in Figure~\ref{SMA}(a). Annex K, bounds-checking interfaces perform bounds checking and call a runtime-constraint handler to deal with the out-of-bounds memory access, usually triggering an exception and terminating the program as presented in Figure~\ref{SMA}(b). In contrast, Figure~\ref{SMA}(c) illustrates how S3Library performs Saturation Memory Access and simply manufactures a boundless memory block (the last padding byte) to the programs as the result of the write operation. The last padding byte always exists that the MinFat allocator guarantees. The implementation of S3Library will be discussed in more detail in Section~\ref{runtime-details}.

\section{Implementation}
\subsection{Overview}
\begin{figure}[ht]
	\setlength{\abovecaptionskip}{0pt}
	\begin{center}
		\includegraphics[width=0.49\textwidth]{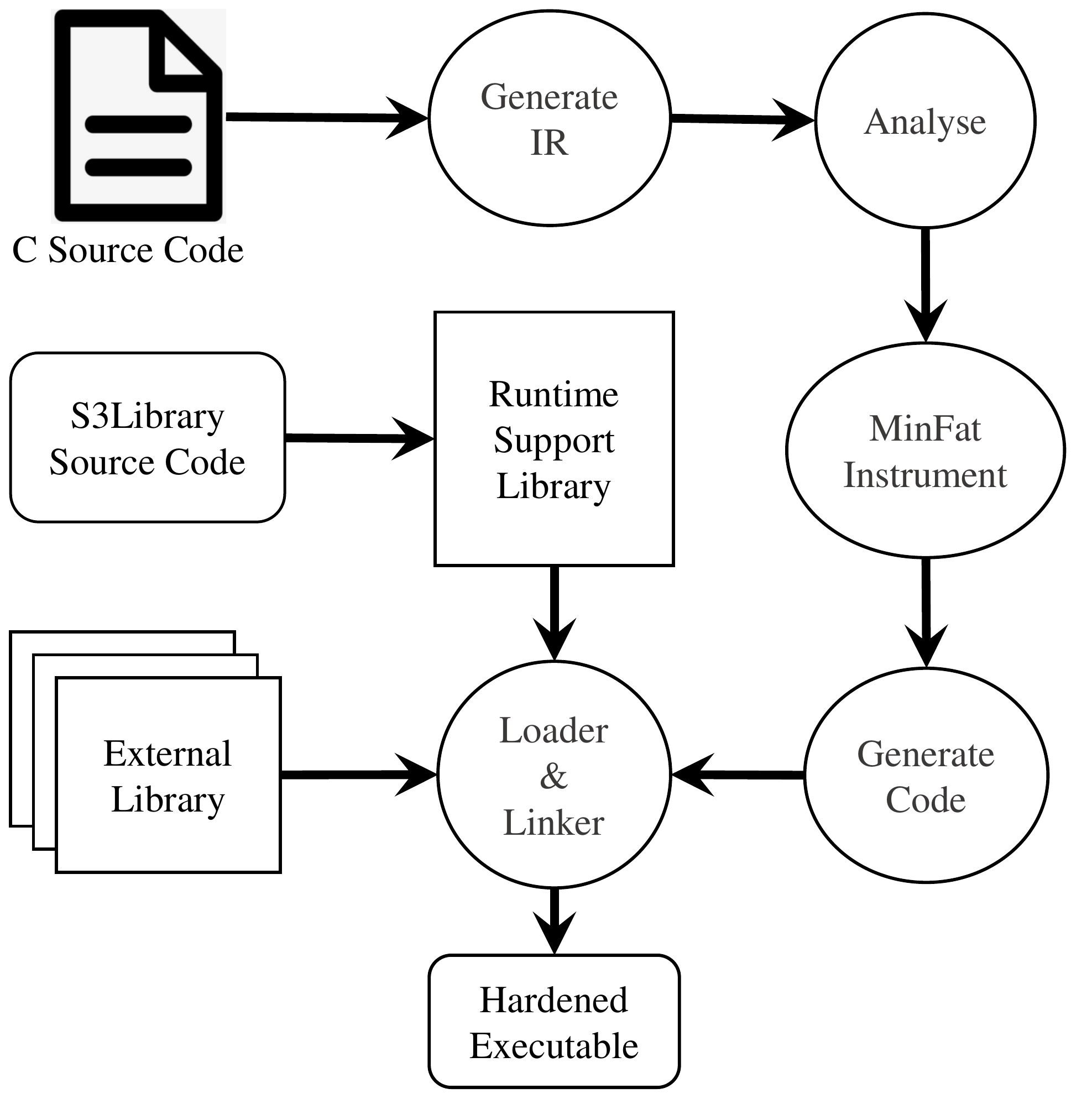}	
	\end{center}
	\caption{\label{overview}Overall system architecture of MinFat and S3Library}
\end{figure}

Figure~\ref{overview} gives an overview of our system architecture.
	
The hardening is performed by a compile-time instrumentation pass. It converts source code to an intermediate representation (IR), generates boundary metadata and replaces the legacy functions name with \_ss suffix. We modify the S3Library source code by inserting the tag-masking operations and Saturation Memory Access to ensure the continued execution rather than triggering a fault or an exception. Then, S3Library source codes are compiled into binary libraries with which the generated code is linked to create a harden executable.
	
We have implemented a prototype of MinFat Pointers and
runtime dynamically loadable S3Library as an extension to clang/LLVM 4.0 for Linux on the x86-64 architecture on top of the LLVM compiler infrastructure. The code consists of 3,797 SLOC of LLVM C++ pass, which adds the instrumentation of MinFat. An additional 5,347 SLOC
make up external libc wrapper (exclude string and memory functions).

\subsection{MinFat implementation}
Our MinFat implementation is based on the approach
LowFat~\cite{duck2017stack,kwon2013low}. MinFat follows the principle of $Less$ $is$ $More$ and adopts a very compact representation for bounds information inspired by BaggyBounds. 

MinFat stack and heap allocators are all implemented as an LLVM compiler infrastructure pass that replaces the default allocation (as represented by the LLVM intermedia representation $alloca$ and $malloc$ instructions). Upon the allocation, all objects are enforced the allocation boundary instead of object boundary, padding and aligning to the nearest power of two. Meanwhile, we store the binary logarithm of the allocation size into the upper six bits within each pointer. Besides, all global variables are all tagged using LLVM pass. Before pointer dereferences, the pointers with TAG have to be masked to create the regular untagged pointer that the CPU expects. Since we focus on buffer overflows resulting from highly-spirit string/memory functions, no checking is performed before every memory access (but addable). For pointer assignment, the result pointer inherits the same TAG of the original. Moreover, we remove the TAG to guarantee fairness when it comes to pointer comparison.

In order to make it easy to use for programmers, we provide an additional compiler flag to invoke our passes during the compilation, just like the way of Address Sanitizer. 
	
\subsection{Runtime Library Support}
\label{runtime-details}

\noindent\textbf{Implementation.} 	
We introduce alternatives for selective unsafe deprecated memory/string, stdio, stdlib and time functions, as listed in Table~\ref{table_S3Library}.
These unsafe functions include almost all C11 Annex K secure versions except <$wchar.h$>. The work of adding wide char functions will be finished in the future. We choose them for the reason that they are the main threats to overflow vulnerability based on experience with real-world security bugs~\cite{howard2006security}. 
In terms of function signatures, all S3Library function names have $\_ss$ suffix, like adding $\_s$ suffix for C11 Annex K functions. The additional “$s$” means the feature of Saturation Memory Access.

\begin{table}[h]  
	\setlength{\abovecaptionskip}{0pt}
	\caption{List of functions in S3Library.} 
	\vspace{-0.2cm}  
	\begin{center}  
		\label{table_S3Library}
		\begin{tabular}{|c|p{6.1cm}|}  
			\hline  
			Header & Functions Name \\ 	
			\hline 
			\hline 
			<$string.h$> & strcpy\_ss, strcat\_ss, strnlen\_ss, strncpy\_ss, strncat\_ss, strtok\_ss, strerror\_ss, memcpy\_ss, memset\_ss, memmove\_ss\\ 
			\hline  
			<$stdio.h$> & gets\_ss, fopen\_ss, freopen\_ss, printf\_ss, fprintf\_ss, vprintf\_ss, vfprintf\_ss, sprintf\_ss, snprintf\_ss, vsprintf\_ss, vsnprintf\_ss, tmpfile\_ss, tmpnam\_ss\\  
			\hline  
			<$stdlib.h$> & getenv\_ss \\
			\hline 
			<$time.h$> & asctime\_ss, ctime\_ss, gmtime\_ss, localtime\_ss \\
			\hline
		\end{tabular}  
	\end{center} 
	\vspace{-0.5cm}  
\end{table}

S3Library is implemented as a dynamically loadable library that must be preloaded for every process to be protected. We believe the fact that runtime checks to verify lengths should be done inside the library functions rather than $wrappers$~\cite{baratloo1999libsafe,avijit2004tied}.
Bounds-checking interfaces perform more checking implementations than what is done in wrappers. For example, NULL pointer checking, zero-length checking, buffer-overlapping checking. If all these
checks are handled with wrappers, it frequently duplicates the work done insides the safe C library.
Consequently, we take away wrappers for these selected string/memory functions, and the program calls their
corresponding $\_ss$ safer versions directly. 

\begin{figure}[t]
	\setlength{\abovecaptionskip}{0pt}
	\begin{center}
	\includegraphics[width=0.5\textwidth]{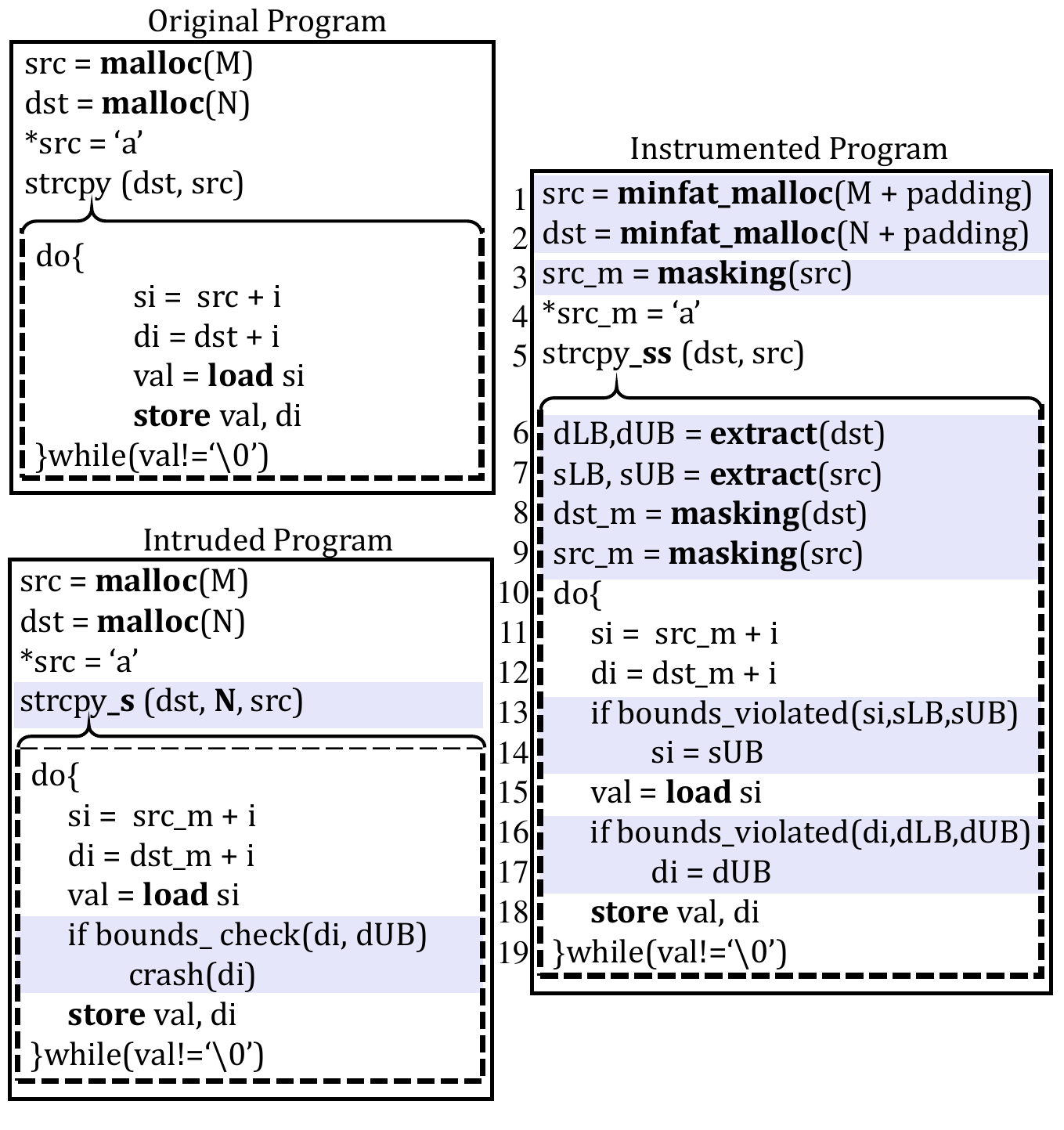}	
	\end{center}
	\caption{\label{code} Memory Safety enforcement of strcpy function code in original program(a) via: C11 Annex K(b) and S3Library(c).}
\end{figure}
	
\noindent\textbf{Illustration.} To illustrate how S3Library works in practice, consider an example in Figure~\ref{code}. Take the strcpy() in an original program as an example. In order to call a bounds-checking interface strcpy\_s, programmers have to manually extend the parameter list with destination buffer size $N$, resulting in an intruded program. S3Library instruments the program and automatically replaces the strcpy with strcpy\_ss. In application codes, heap objects bounds are stored in upper bits for future checks by MINFAT allocator ($minfat\_malloc$) (line 1-2). Pointers with tags have to be explicitly masked before every load/store instruction (line 3) and directly passed into the strcpy\_ss as the parameter with encoded metadata (line 4). At the entry of the strcpy\_ss(), we insert the extracting operations to decode the metadata information (i.e., sLB, dLB, sUB, dUB), and masking operations to mask the tag bits for the following dereferences (line 6-9). Then we verify if the accessed address is within the bound of the referent object on each memory access (line 13,16). In case the runtime-constraints are violated, we perform SMA and manufacture the last padding value to the program as the result of memory read/write (line 14,17), and the instrumented program continues to run without interruption.

\noindent\textbf{Optimization.}
By default, S3Library ensures protection from destination buffer overwrites and source buffer overreads. C standard semantics guarantee that a legal and uncorrupted string variable ends up with `$\backslash0$', thus we can simply return early and finish the copy of string once the source buffer value is `$\backslash0$'. In this way, we can protect only the destination buffer overwrites in S3Library implementation that can already supply sufficiently high security guarantees. As for SMA implementation, all overwrites are directed into the last boundless byte. We can only handle the last overwrite operation using SMA and ignore the other overwrites because the results of all other overwrites are overlapped by the last one.
	
\section{Evaluation}
To evaluate the usefulness and value of our S3Library, we answer the following four research questions:

\noindent\textbf{Q1. Correctness.} Does our S3Library work? Does it break the original work? Does our approach scale to large programs?

\noindent\textbf{Q2. Security.} What level of security is achieved by S3Library according to some security benchmarks like RIPE and NIST? More precisely, does it eliminate buffer overflow vulnerabilities originating from unsafe library function’s call?

\noindent\textbf{Q3. Performance.} What would be the trade-off between security and performance if we only protect highly-critical memory/string functions using a tagged-pointer approach. What is the performance overhead of S3Library in contrast with bounds-checking interfaces?

\noindent\textbf{Q4. Availability.} Does our approach normally run in real-world applications, such as Nginx web server?
	
We made related experiments in order to answer the four questions better. For correctness, we applied our program transformations and safer library on a set of local test suites and SPEC CPU2006 benchmark. For security, we employed the RIPE security benchmark~\cite{wilander2011ripe} and NIST’s SARD reference dataset (Juliet Test Suite for C/C++)~\cite{NIST}. For runtime performance, we first measured overhead at a function call level. Next, the overall performance of C and C++ programs of the SPEC CPU2006 benchmark suite were evaluated. To test availability and usability, we compiled and ran Ngnix Web Server to evaluate our effects on its GET/POST service.

We ran our benchmarks on Intel Xeon E5-2609 machines with 16 cores at 1.70GHz and 32GB of memory, running the 64-bit Ubuntu 16.04.4 LTS. All the benchmarks are compiler with our modified clang version 4.0 with ``-O2 -fanitize=lowfat'' flags and ``-lsafestring-s'' flags to link our S3Library.
	
\subsection{Correctness}
	
We first evaluated the correctness of S3Library using C/C++ programs of SPEC CPU2006. We use the ref input sets and run to completion. To obtain a quantitative estimation on the amounts of instrumented library functions, we utilize ltrace~\cite{branco2007ltrace} to dynamically collect the library information. Table~\ref{table_amounts} presents the histogram of the amounts of the instrumented calls while running SPEC CPU2006. 

\begin{table}[h]  
	\setlength{\abovecaptionskip}{0pt}
	\caption{Numbers of dynamic instrumented calls in SPEC.} 
	\vspace{-0.2cm}  
	\begin{center}  
		\label{table_amounts}
		\scalebox{0.9}{
		\begin{tabular}{|c|c||c|c|}  
			\hline
			$Benchmark$ & $Number$ & $Benchmark$ & $Number$\\
			\hline\hline  
			$401.bzip2$ & 13 & $462.libquantum$ & 8\\
			\hline
			$429.mcf$ & 25,168 & $464.h264ref$ & 222\\
			\hline
			$433.milc$ & 689 & $470.lbm$ & 97 \\
			\hline
			$444.namd$ & 2,909 & $473.astar$ & 27\\
			\hline
			$456.hmmer$ & 31,447 & $482.sphinx3$ & 655,538\\
			\hline
			$458.sjeng$ & 12,779 & &\\
			\hline
		\end{tabular}} 
	\end{center} 
	\vspace{-0.5cm}  
\end{table}
	
Table~\ref{table_amounts} reveals that the number of instrumented calls varies in different benchmarks. For example, a large amount of $strtok$ in 456.hmmer, a great deal of $strlen$ in 482.sphinx3. However, there are almost none instances of functions like memcmp, strerror and time-related calls in SPEC CPU2006, so we make another local test suites for every function to verify the original functionality. The result shows that our implementation passed test sets all.

The running results on SPEC CPU2006 show that our
MinFat approach can pass 7 CINT benchmarks and 3 CFP
benchmarks. Some other benchmarks will malfunction when compiled with MinFat approach because the tagged pointer scheme has inherent compatibility issues. For example, programs like $gcc$ manipulate high bits of pointers, causing damage to the original program. Patches to solve the compatibility issue will be done in future work.

Based on the above validation steps, we conclude that our MinFat and S3Library implementation is correct and valid without impairing the original functionality, thus being able to serve as a solid basis to answer the Q1.
	
\begin{figure*}[h]
	\setlength{\abovecaptionskip}{-5pt}
	\begin{center}
		\includegraphics[width=1.0\textwidth,height=0.25\textheight]{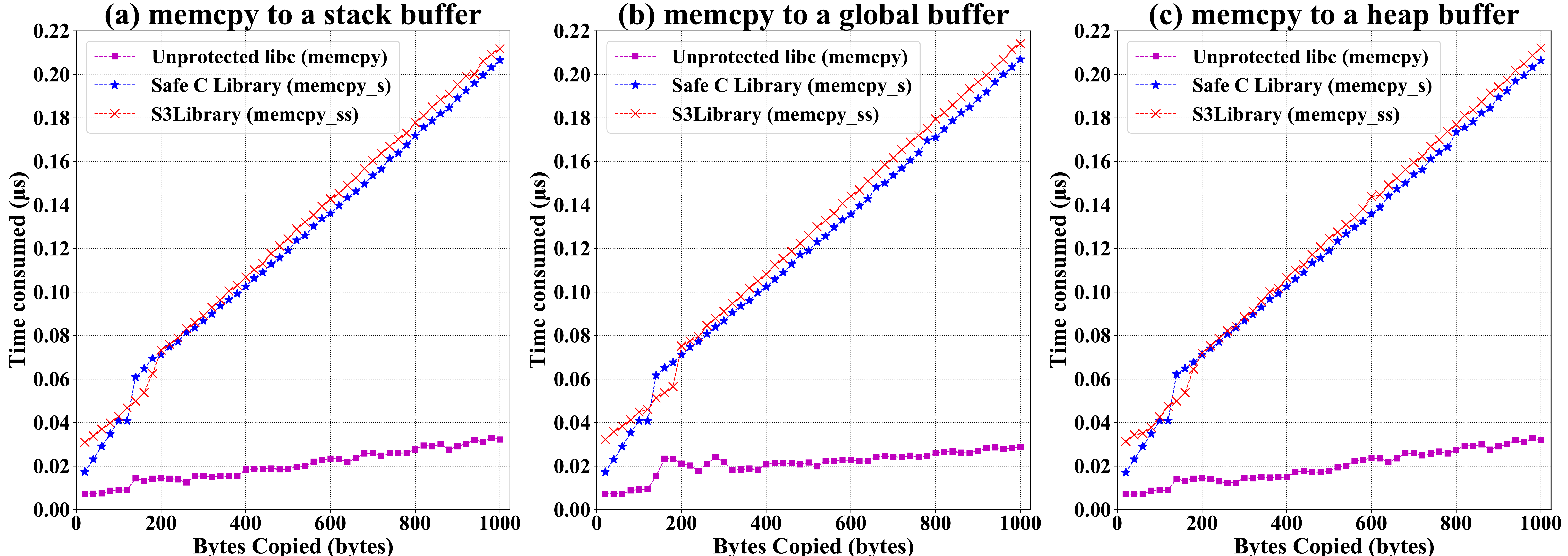}		
	\end{center}
	\caption{\label{memcpy} Memcpy to a stack, global and heap buffer.}
\end{figure*}	
	
\subsection{Security}
To test the security guarantees of our S3Library, we employed the RIPE benchmark~\cite{wilander2011ripe} and NIST’s SARD reference dataset (Juliet Test Suite for C/C++)~\cite{NIST}. The results are shown in Table~\ref{table_RIPE} and Table~\ref{table_NIST}.

\noindent\textbf{RIPE.} An extension of Wilander’s Lamkar’s security benchmark, provides a standard way of testing the coverage of a defense mechanism against buffer overflow.
RIPE claims to cover 850 working buffer overflow attacks from five dimensions including location, target code pointer, overflow technique, attack code and function abused.

In order to make native testings on RIPE benchmarks, we firstly deployed a 64-bit RIPE because the original RIPE only supports 32-bit. Secondly, we removed 4 sub-object overflows attacks and ``homebrew'' function attack (a loop-based equivalent of memcpy in non-library code) because they are outside the protection boundary of our implementation.

	
We tested all the remaining RIPE parameter combinations (735 valid attacks in total). Table~\ref{table_RIPE} shows the security results of all approaches. Successfully means arbitrary code execution. Failed attacks are repeatably prevented. GCC 5.4.0 -O2 could not detect 33 out of these 735 attacks and the number of MinFat+glibc is 2, which were all about abusing “memcpy”. Our MinFat+S3Library implementation prevented the remaining two $memcpy$ attacks. One attack was eliminated by Saturation Memory Access within the $memcpy\_ss$ function. The other one was inhibited by the length check at the entry, invoking an exception and returning in advance. The result shows that all attacks had been successfully prevented by our MinFat and S3Library implementation.

\begin{table}[h]  
	\setlength{\abovecaptionskip}{0pt}
	\caption{Results of RIPE security benchmark.} 
	\vspace{-0.2cm}  
	\begin{center}  
		\label{table_RIPE}
		\scalebox{0.9}{
			\begin{tabular}{|c||c|c|c|}  
			\hline
			Implementation & Successfully & Failed  & Total\\
			\hline\hline
			$GCC 5.4.0$ & 33 & 702 & 735\\
			\hline
			$MinFat$+$glibc$ & 2 & 733 & 735\\
			\hline
			$MinFat$+$S3Library$ & 0 & 735 & 735\\
			\hline
		\end{tabular}} 
	\end{center} 
	\vspace{-0.5cm}  
\end{table}

\noindent\textbf{NIST.} NIST's SARD (Software Assurance Reference Dataset) is a collection of thousands of test programs with known security flaws. SARD’s Juliet C/C++ Test Suite version 1.3 provides 64,099 test cases and more than 100,000 files, which is referred to as the most comprehensive benchmark
available for C/C++ buffer overflow vulnerabilities. We select CWE121: Stack-based Buffer Overflow and CWE-122: Heap-based Buffer Overflow for the experiment. These two test suites are related to unsafe library functions calling.

We mainly tested three most common vulnerable functions: $memcpy$, $strcpy$ and $strcat$, which we consider are the main threats to buffer overflow flaws in a large amount of real-life open-source programs. S3Library was applied to 1,939
programs across CWE121 and CWE122. Table~\ref{table_NIST} shows the
amounts of vulnerable programs and involved unsafe functions we have tested.

CWE 121 consists of 854 programs representing stack-based buffer overflow caused by abusing memcpy, strcpy and strcat. In these programs, a fixed-size dest buffer is created in stack memory space and then assigned with values from src buffer that is too large for it to hold. Our implementation first reallocates pow-of-two objects for them. These boundless padding blocks, to some extend, tolerates the difference of size between src and dest as long as their size is identical after
pow-of-two alignment. Finally, S3Library, such as memcpy\_ss,  will perform bounds checking and SMA to eliminate out-of-bound memory access whenever any runtime-constraint occurs.

CWE 122 consists of 1,085 programs representing heap-based buffer overflow caused by abusing memcpy, strcpy and strcat. The difference between CWE 122 and CWE 121 that the dest buffer is allocated in heap memory. MinFat+S3Library executes a similar behavior. MinFat tags the pointers, replaces unsafe functions with suffix $\_ss$ and calls the $\_ss$ functions at runtime.
	
Program in Juliet C/C++ version 1.3 has a good function and a bad function in each file. The good function passes the correct and appropriate size to the function to perform some string or memory operations. The bad function passes a wrong
and inappropriate size and produces either a segment fault or incorrect output results. Teble~\ref{table_NIST} presents the result that our 
MinFat and S3Library implementation has eliminated all 1,939 vulnerabilities in bad functions.

\begin{table}[h]  
	\setlength{\abovecaptionskip}{0pt}
	\caption{Results of NIST’s Juliet C/C++ Test Suites.} 
	\vspace{-0.2cm}  
	\begin{center}  
		\label{table_NIST}
		\scalebox{0.9}{
			\begin{tabular}{|c||c|c|c|}  
			\hline
			Buffer Overflow Type& $memcpy$ & $strcpy$  & $strcat$\\
			\hline\hline
			$CWE$ 121: $Stack$-$based$ & 434 &248 &172\\
			\hline
			$CWE$ 122: $Heap$-$based$ & 624 &269 &192\\
			\hline
			$Total$ &1058 &517& 364\\
			\hline
		\end{tabular}}  
	\end{center} 
	\vspace{-0.5cm}  
\end{table}

Based on the above validation steps, we conclude that our MinFat and S3Library implementation is security enough that original programs run normally without segment fault and buffer overflow followed by a control flow hijack did not occur, serving as a solid basis to answer the Q2.
	
\begin{figure*}[t]
	\setlength{\abovecaptionskip}{-5pt}
	\begin{center}
		\includegraphics[width=1.0\textwidth]{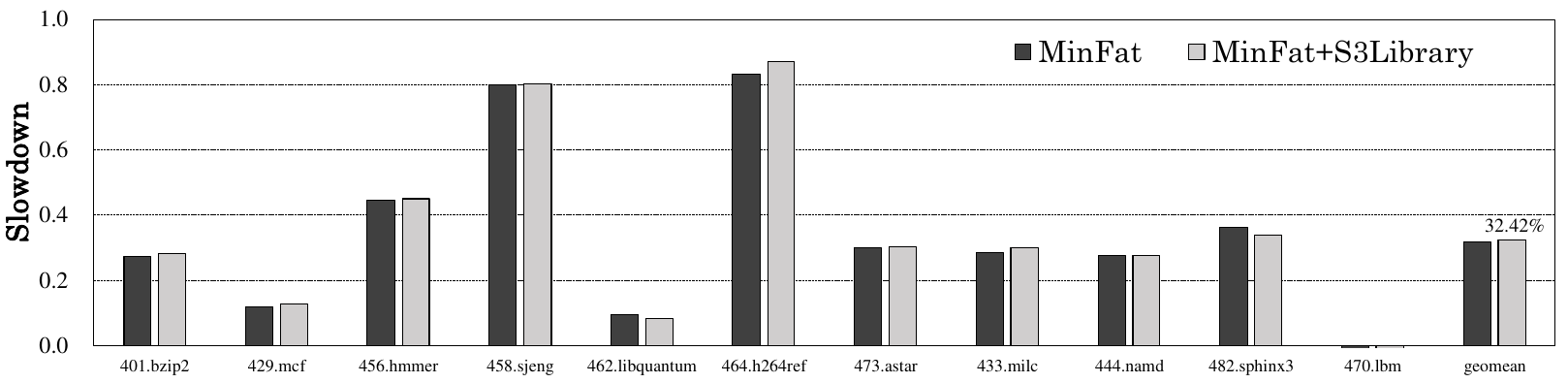}		
	\end{center}
	\caption{\label{time} Performance Overhead on SPEC 2006}
	 \vspace{-0.5cm} 
\end{figure*}
		
\subsection{Runtime Performance}
\label{runtime overhead}

\noindent\textbf{Microbenchmark.}
First, we need to figure out the overhead at a function call level. We present the comparison of execution times of one most commonly used function: $memcpy()$ for the following three cases. 
\begin{itemize}
	\item[-] $Legacy$ function without any protection.
	\item[-] Safe $C$ $Library$ using implementations of C11 Annex K, bounds-checking interfaces.
	\item[-] $S3Library$ with the same API as legacy function but performs bounds checking.
\end{itemize}

The time required by a single $memcpy()$ from the same local buffer into global, stack and heap buffer was measured for varying numbers of bytes copied. We
use “-fno-builtin-” flags to disable the inline optimization in clang/LLVM. As presented in Figure~\ref{memcpy}, there is no significant difference in performance overheads of different types of destination buffers. As we can expect, bounds checking progress leads to several times overhead as the number of bytes copied increases. The performance effects on additional extracting and masking progress in S3Library can be negligible when the number of bytes exceeds 200.

\noindent\textbf{SPEC CPU2006 benchmarks.} Next, we report the overhead over the SPEC CPU2006 benchmark suite. We compare three versions of implementations:

\begin{itemize}
	\item[-]  $Base$: The uninstrumented program that uses the default allocator and legacy functions.
	\item[-] $MinFat$+$glibc$: The program instrumented with MinFat, but calling standard functions in wrappers.
	\item[-] $MinFat$+$S3Library$: The instrumented program with MinFat and instrumented S3Library calls.
\end{itemize}

We run each benchmark-input pair three times and use the average execution time as its performance.
Figure \ref{time} shows the performance overhead normalized to the baseline (Base). Each overhead number is the average of 3 iterations of the same program (using the reference workset for SPEC). On average, MinFat+S3Lirary incurs a 1.32x slowdown.
	
We observe that the performance difference between MinFat+S3library and MinFat+glibc is negligible. This is expected because the number of call instructions, as shown in Table~\ref{table_amounts}, is much lower than the number of load/store and allocation instructions in SPEC~\cite{phansalkar2007analysis}. Without the checks, tagging and masking operations are the main associated performance overhead caused by pointer-based approaches, especially for pointer-intensive programs. 456.hmmer and 464.h264ref has top 2 load/store instructions ratios, 65.9\% and 54.9\% respectively~\cite{phansalkar2007analysis}. Additionally, 458.sjeng is a program that plays chess and there is a large number of local buffers to be allocated in programs, taking more time for tagging operations. 464.h264ref that is related to video compression has the same situation~\cite{henning2006spec}. In total, those are the reasons why these three programs have relatively higher performance overhead.

We conclude that the implementation of MinFat is much more expensive than S3Library, and tasks on how to optimize the overhead of tagging and masking operations will be discussed in Section~\ref{sec:discussion}.
	
\begin{table}[h]  
	\setlength{\abovecaptionskip}{0pt}
	\caption{Memory Overhead on SPEC 2006.} 
	\vspace{-0.2cm}  
	\begin{center}  
		\label{mem}
		\scalebox{0.9}{
		\begin{tabular}{|c||c|c|}  
			\hline
			$Benchmark$ & $MinFat$+$glibc$  & $MinFat$+$S3Library$\\
			\hline\hline  
			$401.bzip2$ & -0.30\% & -0.33\% \\
			\hline
			$429.mcf$ & 0.03\% &  0.01\%\\
			\hline
			$433.milc$ & 0.13\% & 0.11\% \\
			\hline
			$444.namd$ & 18.86\% & 18.19\% \\
			\hline
			$456.hmmer$ & 40.82\% & 40.16\%\\
			\hline
			$458.sjeng$ & 0.72\% & 0.62\% \\
			\hline
			$462.libquantum$ & 27.08\%& 17.79\%\\
			\hline
			$464.h264ref$&65.07\% & 64.75\%\\
			 \hline
			 $470.lbm$&0.09\%&0.05\%\\
			 \hline$473.astar$&189.72\%&189.88\%\\
			 \hline$482.sphinx3$&63.00\%&62.48\%\\
			 \hline$gomean$ &29.16\%&28.08\%\\
			\hline
		\end{tabular}} 
	\end{center} 
	\vspace{-0.5cm}  
\end{table}


To determine the impact of our scheme on memory usage, we have measured the mean resident set size (RSS) while running the SPEC CPU2006 benchmark suite. Table~\ref{mem} presents the memory overhead on SPEC2006.
MinFat utilizes a straightforward encoding scheme. However, the simple encoding results in obvious memory overhead through fragmentation. For example, 473.astar reads various map region size for region-based path finding algorithm, resulting in large memory fragments. Fortunately, there are 10 unused bits remaining to improve our encoding and reduce internal fragments. This work will be done in the future. Besides, MinFat+S3Library calls directly $\_ss$ functions rather than calling through wrappers, which helps reduce memory overhead.
	
\subsection{Availability}
	
For availability, we tested Nginx version 1.16.1 (the stable version). Nginx is a free, open-source and lightweight HTTP server. To successfully run Nginx under MinFat, we manually add llvm wrappers for each I/O system-call function to mask the argument pointers with tags and insert an uninstrumented statement before each inline assembly. 

The ab benchmark was run on a client machine to generate workload. We test two types of modes: GET and POST. For Get mode, we fetch a static web page from localhost via HTTP. To adapt the load, we set a fixed number of issued requests and increase the amounts of concurrent requests at a time. The results are shown in Figure \ref{fig-nginx}. We observe negligible effects on performance because Nginx is not a primarily CPU-bound application and does relatively few memory operations.

\begin{table}[h] 	
	\setlength{\abovecaptionskip}{0pt}
	\caption{Nginx HTTP server benchmark POST overhead.} 
	
	\vspace{-0.2cm}  
	\begin{center}  
		\label{table-nginx}
		\scalebox{0.9}{
			\begin{tabular}{|l|l|l|}  
				\hline  
				Implementation & Timings(ms) & Memory(KB) \\ 
				\hline
				\hline  
				nginx-1.16.1-$Baseline$ & 121.48 & 414976\\ 
				\hline  
				nginx-1.16.1-$MinFat$ & 128.9(6.11\%) & 417300(0.56\%)\\  
				\hline  
				nginx-1.16.1-$S3Library$ & 129.5(6.60\%) & 417048(0.50\%) \\
				\hline
		\end{tabular}}  
	\end{center} 
	\vspace{-0.5cm}  
\end{table}
	
The implementation of POST mode further confirms the result. For this test, we transfer a relatively small file (200MB) to the local host with the purpose to stress the CPU and provide an objective result using MinFat. Each test was run for a total of 50 times and results were averaged. For the time consumption and memory usage, both overheads of MinFat and S3Library was very low. On average, we observe a 6.60\% increase in latency and a 0.50\% increase in memory bloat, as shown in Table \ref{table-nginx}.
	
\begin{figure}[t]
	\setlength{\abovecaptionskip}{0pt}
    \begin{center}
		\scalebox{0.9}{\includegraphics[width=0.49\textwidth]{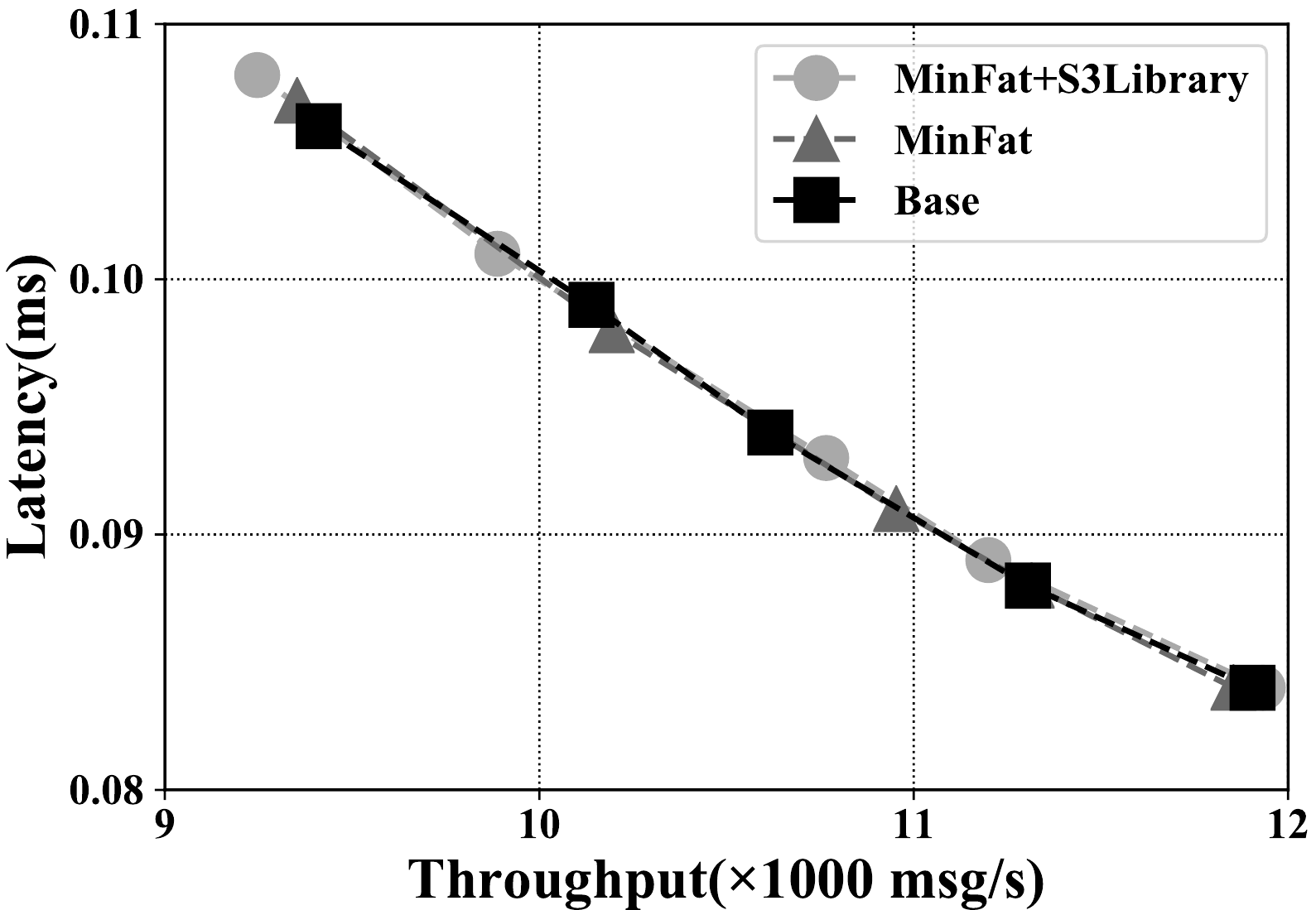}}		
	\end{center}
	\caption{\label{fig-nginx} Throughout-Latency for Nginx web server.}
\end{figure}

\section{Related Work}
	
The Libsafeplus technique~\cite{avijit2004tied} shares some common motivations and ideas with S3Library. However, S3Library performs more checking implementations than bounds checking done in wrappers. For example, buffer-overlapping checking. We think if all these checkings are handled with wrappers, it duplicates the work done inside the safe C library. This is the reason why we remove wrappers for our S3Library. Moreover, Libsafeplus needs the target program compiled with -$g$ option in advance to extract the debugging information, which is not always allowed in many cases.  
	
	
\section{Discussion and Future Work}
\label{sec:discussion}    
\noindent\textbf{Encoding Scheme.} Our MinFat Pointer prototype utilizes 6 bits to address $2^{64}$ bytes virtual space. This is the minimum encoding scheme by default (but configurable) with obvious memory bloat. However, the remaining 10 bits are unused since the upper bit of a 48-bit pointer is reserved for kernel space in most operating systems. Thus efficient allocation algorithm and safer memory safety scheme with these bits will be explored in our future work. 

\noindent\textbf{Performance Overhead.} 
The main runtime overhead results from two parts: tagging on each live object and masking before each memory access. Some architectures, e.g., AArch64, provide hardware supports for virtual address tagging and allow MMU to ignore the upper bits during address translation~\cite{holdingsarm}. Thus the masking operation before each pointer dereference can be omitted. Besides, our scheme can be more effective if MinFat's implementation is supported by instruction set extensions. For example, tag-masking operations can be integrated into the execution of load/store instruction and accelerated by related hardware units.
	
	
	
\noindent\textbf{Coverage.}
Additional S3Library functions also make strong candidates, such as POSIX and other ANSI C. Our approach still has to extend the secure functions list and provides a good coverage, although a thorough examination of industry practice is required to make a complete list. 

\section{Conclusion}
Attacks that exploit buffer overflow vulnerabilities in C and
C++ programs continue to be a severe security problem. In this paper we present S3Library, a safer library reinforcement that automatically replaces deprecated string/memcpy functions with safer versions that perform bounds checking. Once detecting a buffer overflow vulnerability, S3Library eliminates it via boundless padding blocks that MinFat Pointer allocator supplies, rather than terminate the program. Our approach explores the space of how tagged pointer approaches trade security and overhead without the checks.

\bibliographystyle{plain}
\bibliography{\jobname}

\begin{thebibliography}{10}

\bibitem{cwetop25}
Cwe - common weakness enumeration.
\newblock \url{https://cwe.mitre.org/index.html}.

\bibitem{IntelSafeString}
Intel safe string library.
\newblock \url{https://github.com/intel/safestringlib}.

\bibitem{twitter}
Matt miller. 2017. heap corruption issues reported to microsoft. (2017).
\newblock \url{https://twitter.com/epakskape/status/851479629873332224}.

\bibitem{NIST}
Nist software assurance reference dataset project.
\newblock \url{https://samate.nist.gov/SARD/}.

\bibitem{openwatcom}
Open watcom c++ class library reference version 1.8.
\newblock \url{ftp://ftp.openwatcom.org/manuals/current/cpplib.pdf}.

\bibitem{SafeCLibrary}
Safe c library - iso tr24731 bounds checking interface.
\newblock \url{https://github.com/rurban/safeclib/}.

\bibitem{akritidis2009baggy}
Periklis Akritidis, Manuel Costa, Miguel Castro, and Steven Hand.
\newblock Baggy bounds checking: An efficient and backwards-compatible defense
  against out-of-bounds errors.
\newblock In {\em USENIX Security Symposium}, pages 51--66, 2009.

\bibitem{avijit2004tied}
Kumar Avijit, Prateek Gupta, and Deepak Gupta.
\newblock Tied, libsafeplus: Tools for runtime buffer overflow protection.
\newblock In {\em USENIX Security Symposium}, pages 45--56, 2004.

\bibitem{baratloo1999libsafe}
Arash Baratloo, Navjot Singh, and Timothy Tsai.
\newblock Libsafe: Protecting critical elements of stacks.
\newblock {\em White Paper http://www. research. avayalabs.
  com/project/libsafe}, 1999.

\bibitem{branco2007ltrace}
Rodrigo~Rubira Branco.
\newblock Ltrace internals.
\newblock In {\em Proceedings of the Linux Symposium}, volume~1, pages 41--52.
  Ottawa, ON, Canada, June, 2007.

\bibitem{brunink2011boundless}
Marc Br{\"u}nink, Martin S{\"u}{\ss}kraut, and Christof Fetzer.
\newblock Boundless memory allocations for memory safety and high availability.
\newblock In {\em 2011 IEEE/IFIP 41st International Conference on Dependable
  Systems \& Networks (DSN)}, pages 13--24. IEEE, 2011.

\bibitem{N1967}
Martin~Sebor Carlos~O'Donell.
\newblock Field experience with annex k — bounds checking interfaces.
\newblock \url{http://www.open-std.org/jtc1/sc22/wg14/www/docs/n1967.htm/},
  September 25, 2015.
\newblock Accessed 30 March, 2020.

\bibitem{duck2017stack}
Gregory~J Duck, Roland~HC Yap, and Lorenzo Cavallaro.
\newblock Stack bounds protection with low fat pointers.
\newblock In {\em NDSS}, 2017.

\bibitem{henning2006spec}
John~L Henning.
\newblock Spec cpu2006 benchmark descriptions.
\newblock {\em ACM SIGARCH Computer Architecture News}, 34(4):1--17, 2006.

\bibitem{holdingsarm}
Arm Holdings.
\newblock Arm cortex-a series programmer’s guide for armv8-a-15.2. dynamic
  voltage and frequency scaling.

\bibitem{homaei2017seven}
Hossein Homaei and Hamid~Reza Shahriari.
\newblock Seven years of software vulnerabilities: The ebb and flow.
\newblock {\em IEEE Security \& Privacy}, 15(1):58--65, 2017.

\bibitem{howard2006security}
Michael Howard and Steve Lipner.
\newblock {\em The security development lifecycle}, volume~8.
\newblock Microsoft Press Redmond, 2006.

\bibitem{jones1997backwards}
Richard~WM Jones and Paul~HJ Kelly.
\newblock Backwards-compatible bounds checking for arrays and pointers in c
  programs.
\newblock In {\em Proceedings of the 3rd International Workshop on Automatic
  Debugging; 1997 (AADEBUG-97)}, number 001, pages 13--26. Link{\"o}ping
  University Electronic Press, 1997.

\bibitem{jtc2011sc22}
ISO Jtc.
\newblock Sc22/wg14. iso/iec 9899: 2011.
\newblock {\em Information technology—Programming languages—C. http://www.
  iso. org/iso/iso\_catalogue/catalogue\_ tc/catalogue\_detail. htm}, 2011.

\bibitem{kroes2017fast}
Taddeus Kroes, Koen Koning, Cristiano Giuffrida, Herbert Bos, and Erik van~der
  Kouwe.
\newblock Fast and generic metadata management with mid-fat pointers.
\newblock In {\em Proceedings of the 10th European Workshop on Systems
  Security}, pages 1--6, 2017.

\bibitem{kroes2018delta}
Taddeus Kroes, Koen Koning, Erik van~der Kouwe, Herbert Bos, and Cristiano
  Giuffrida.
\newblock Delta pointers: Buffer overflow checks without the checks.
\newblock In {\em Proceedings of the Thirteenth EuroSys Conference}, pages
  1--14, 2018.

\bibitem{kuvaiskii2017sgxbounds}
Dmitrii Kuvaiskii, Oleksii Oleksenko, Sergei Arnautov, Bohdan Trach, Pramod
  Bhatotia, Pascal Felber, and Christof Fetzer.
\newblock Sgxbounds: Memory safety for shielded execution.
\newblock In {\em Proceedings of the Twelfth European Conference on Computer
  Systems}, pages 205--221, 2017.

\bibitem{kwon2013low}
Albert Kwon, Udit Dhawan, Jonathan~M Smith, Thomas~F Knight~Jr, and Andre
  DeHon.
\newblock Low-fat pointers: compact encoding and efficient gate-level
  implementation of fat pointers for spatial safety and capability-based
  security.
\newblock In {\em Proceedings of the 2013 ACM SIGSAC conference on Computer \&
  communications security}, pages 721--732, 2013.

\bibitem{laverdiere2009implementation}
Marc-Andr{\'e} Laverdi{\`e}re, Serguei~A Mokhov, and Djamel Benredjem.
\newblock On implementation of a safer c library, iso/iec tr 24731.
\newblock {\em arXiv preprint arXiv:0906.2512}, 2009.

\bibitem{monperrus2018automatic}
Martin Monperrus.
\newblock Automatic software repair: a bibliography.
\newblock {\em ACM Computing Surveys (CSUR)}, 51(1):1--24, 2018.

\bibitem{N1106}
WG14 N1106.
\newblock N1106 austin group review of iso/iec wdtr 24731 specification for
  secure c library functions.
\newblock \url{http://www.open-std.org/jtc1/sc22/ wg14/www/docs/n1106.txt/},
  March 07, 2005.
\newblock Accessed 30 March, 2020.

\bibitem{nagarakatte2015everything}
Santosh Nagarakatte, Milo~MK Martin, and Steve Zdancewic.
\newblock Everything you want to know about pointer-based checking.
\newblock In {\em 1st Summit on Advances in Programming Languages (SNAPL
  2015)}. Schloss Dagstuhl-Leibniz-Zentrum fuer Informatik, 2015.

\bibitem{nagarakatte2009softbound}
Santosh Nagarakatte, Jianzhou Zhao, Milo~MK Martin, and Steve Zdancewic.
\newblock Softbound: Highly compatible and complete spatial memory safety for
  c.
\newblock In {\em Proceedings of the 30th ACM SIGPLAN Conference on Programming
  Language Design and Implementation}, pages 245--258, 2009.

\bibitem{nagarakatte2010cets}
Santosh Nagarakatte, Jianzhou Zhao, Milo~MK Martin, and Steve Zdancewic.
\newblock Cets: compiler enforced temporal safety for c.
\newblock In {\em Proceedings of the 2010 international symposium on Memory
  management}, pages 31--40, 2010.

\bibitem{oleksenko2017intel}
Oleksii Oleksenko, Dmitrii Kuvaiskii, Pramod Bhatotia, Pascal Felber, and
  Christof Fetzer.
\newblock Intel mpx explained: An empirical study of intel mpx and
  software-based bounds checking approaches.
\newblock {\em arXiv preprint arXiv:1702.00719}, 2017.

\bibitem{phansalkar2007analysis}
Aashish Phansalkar, Ajay Joshi, and Lizy~K John.
\newblock Analysis of redundancy and application balance in the spec cpu2006
  benchmark suite.
\newblock In {\em Proceedings of the 34th annual international symposium on
  Computer architecture}, pages 412--423, 2007.

\bibitem{rinard2003acceptability}
Martin Rinard.
\newblock Acceptability-oriented computing.
\newblock {\em Acm sigplan notices}, 38(12):57--75, 2003.

\bibitem{rinard2004dynamic}
Martin Rinard, Cristian Cadar, Daniel Dumitran, Daniel~M Roy, and Tudor Leu.
\newblock A dynamic technique for eliminating buffer overflow vulnerabilities
  (and other memory errors).
\newblock In {\em 20th Annual Computer Security Applications Conference}, pages
  82--90. IEEE, 2004.

\bibitem{rinard2004enhancing}
Martin~C Rinard, Cristian Cadar, Daniel Dumitran, Daniel~M Roy, Tudor Leu, and
  William~S Beebee.
\newblock Enhancing server availability and security through failure-oblivious
  computing.
\newblock In {\em OSDI}, volume~4, pages 21--21, 2004.

\bibitem{ruwase2004practical}
Olatunji Ruwase and Monica~S Lam.
\newblock A practical dynamic buffer overflow detector.
\newblock In {\em NDSS}, volume 2004, pages 159--169, 2004.

\bibitem{SLIBC}
Austria sba research.
\newblock Implementation of c11 annex k "bounds-checking interfaces" iso/iec
  9899:2011.
\newblock \url{https://code.google.com/archive/p/slibc/}, 2012.

\bibitem{N997}
SC22/WG14/N997.
\newblock Proposal for technical report on c standard library security.
\newblock Technical report, February 24, 2003.

\bibitem{seacordbounds}
Robert~C Seacord.
\newblock Bounds-checking interfaces: Field experience and future directions.

\bibitem{serebryany2012addresssanitizer}
Konstantin Serebryany, Derek Bruening, Alexander Potapenko, and Dmitriy Vyukov.
\newblock Addresssanitizer: A fast address sanity checker.
\newblock In {\em Presented as part of the 2012 $\{$USENIX$\}$ Annual Technical
  Conference ($\{$USENIX$\}$$\{$ATC$\}$ 12)}, pages 309--318, 2012.

\bibitem{serebryany2018memory}
Kostya Serebryany, Evgenii Stepanov, Aleksey Shlyapnikov, Vlad Tsyrklevich, and
  Dmitry Vyukov.
\newblock Memory tagging and how it improves c/c++ memory safety.
\newblock {\em arXiv preprint arXiv:1802.09517}, 2018.

\bibitem{shahriar2013buffer}
Hossain Shahriar, Hisham~M Haddad, and Ishan Vaidya.
\newblock Buffer overflow patching for c and c++ programs: rule-based approach.
\newblock {\em ACM SIGAPP Applied Computing Review}, 13(2):8--19, 2013.

\bibitem{shaw2014automatically}
Alex Shaw, Dusten Doggett, and Munawar Hafiz.
\newblock Automatically fixing c buffer overflows using program
  transformations.
\newblock In {\em 2014 44th Annual IEEE/IFIP International Conference on
  Dependable Systems and Networks}, pages 124--135. IEEE, 2014.

\bibitem{smirnov2007automatic}
Alexey Smirnov and Tzi-cker Chiueh.
\newblock Automatic patch generation for buffer overflow attacks.
\newblock In {\em Third International Symposium on Information Assurance and
  Security}, pages 165--170. IEEE, 2007.

\bibitem{szekeres2013sok}
Laszlo Szekeres, Mathias Payer, Tao Wei, and Dawn Song.
\newblock Sok: Eternal war in memory.
\newblock In {\em 2013 IEEE Symposium on Security and Privacy}, pages 48--62.
  IEEE, 2013.

\bibitem{TR24731}
ISO/IEC JTC1~SC22 WG14.
\newblock Information technology, programming languages, their environments and
  system software interfaces, extensions to the c library, part i:
  Bounds-checking interfaces.
\newblock Technical report, March 28, 2007.

\bibitem{wilander2011ripe}
John Wilander, Nick Nikiforakis, Yves Younan, Mariam Kamkar, and Wouter Joosen.
\newblock Ripe: runtime intrusion prevention evaluator.
\newblock In {\em Proceedings of the 27th Annual Computer Security Applications
  Conference}, pages 41--50, 2011.

\bibitem{woodruff2014cheri}
Jonathan Woodruff, Robert~NM Watson, David Chisnall, Simon~W Moore, Jonathan
  Anderson, Brooks Davis, Ben Laurie, Peter~G Neumann, Robert Norton, and
  Michael Roe.
\newblock The cheri capability model: Revisiting risc in an age of risk.
\newblock In {\em 2014 ACM/IEEE 41st International Symposium on Computer
  Architecture (ISCA)}, pages 457--468. IEEE, 2014.

\bibitem{xu2004efficient}
Wei Xu, Daniel~C DuVarney, and R~Sekar.
\newblock An efficient and backwards-compatible transformation to ensure memory
  safety of c programs.
\newblock In {\em Proceedings of the 12th ACM SIGSOFT twelfth international
  symposium on Foundations of software engineering}, pages 117--126, 2004.

\end{thebibliography}
	

\end{document}